\documentstyle[prl,aps,twocolumn]{revtex} 

\newcommand{\r}{{\bf r}}

\begin{document}
\draft
\title{Comparison of mean-field theories for
vortices in trapped Bose-Einstein condensates}

\author{S M M Virtanen, T P Simula and M M Salomaa}
\address{Materials Physics Laboratory, 
Helsinki University of Technology\\
P.~O.~Box 2200 (Technical Physics), FIN-02015 HUT, Finland}
\date{\today}

\maketitle
\begin{abstract}
We compute structures of vortex configurations in a harmonically
trapped Bose-Einstein condensed atom gas within three different
gapless self-consistent mean-field theories. Outside the vortex
core region, the density profiles for the condensate and the thermal gas
are found to differ only by a few percent between the
Hartree-Fock-Bogoliubov-Popov theory and two of its recently
proposed gapless extensions. In the core region, however, the
differences in the density profiles are substantial.
The structural differences are reflected in the energies of
the quasiparticle states localized near the vortex core. 
Especially, the predictions for the energy of the lowest 
quasiparticle excitation differ considerably between the theoretical
models investigated. 
\end{abstract}
\hspace{5mm}
\pacs{PACS number(s): 03.75.Fi, 05.30.Jp, 67.40.Db}

%\narrowtext
The landmark experiments to realize Bose-Einstein condensation
in dilute atomic gases \cite{first_exp} have sparked vigorous investigation
on the physical properties of these novel quantum fluids. Due to the
weak interactions, such systems are rare examples of interacting
quantum fluids amenable to quantitative microscopic analysis, and
thus provide unique possibilities to test the fundamental principles and 
theories of many-body quantum physics. Theoretical approaches 
yield several quantities, such as density profiles for the
condensate and the thermal gas component, stability estimates,
specific heats, and properties of various propagating sound modes, 
to be compared with experiments. Experiments also yield detailed 
information on the energies of the individual 
excitation modes of these systems \cite{exp_ex}. Such information provides the
most direct and stringent tests for the accuracy of different
theoretical approaches, as compared to the above-mentioned ``collective'' 
quantities which depend on the excitation spectrum as a whole. 

The Bogoliubov equations \cite{bog_fet} are a widely used starting 
point to compute the excitation spectra for dilute Bose-Einstein 
condensates (BECs).
They can be seen as eigenmode equations for the condensate described
by the Gross-Pitaevskii equation \cite{gross_pitaevskii}, 
neglecting effects of the
thermal, noncondensed gas component in the system.
The Hartree-Fock-Bogoliubov (HFB) theory  \cite{griffin_form}
takes self-consistently into account 
the condensate and the thermal gas densities, as well as the lowest
order anomalous average of the boson field. 
However, it is plagued by an unphysical gap in the excitation spectrum,
which violates Goldstone's theorem
and invalidates its value in predicting the lowest
collective mode excitation frequencies. Goldstone's theorem can be restored
by neglecting the anomalous average mean field in the HFB formalism.
This yields the gapless Popov version of the HFB theory \cite{popov}. At low 
temperatures, predictions of the Popov approximation (PA) 
for the lowest excitation frequencies of irrotational 
condensates are in good agreement with experimental 
results, but at temperatures $T\gtrsim T_{\sc bec} / 2$ 
($T_{\sc bec}$ denotes the critical temperature of
condensation) the deviations become apparent \cite{popov_exp}. 
The main inadequacies of the PA are that it neglects the effects of 
the background gas on atomic collisions and the dynamics of
the thermal gas component. As an improvement to overcome 
the first limitation within a computationally manageable formalism, 
the so-called G1 and G2 approximations have been suggested 
\cite{gener1,gener2}. They are gapless mean-field 
theories which take into account effects of the medium
on atomic collisions by allowing the interaction couplings to depend
on the correlation mean fields in a self-consistent manner. 
The two versions are based on different approximations for the
momentum dependence of the full many-body $T$-matrix in the 
homogeneous gas limit, and their precision for inhomogeneous 
systems remains to be investigated.

To assess the accuracy of the above-mentioned gapless HFB-type
approximations, their predictions for the excitation frequencies
of irrotational, harmonically trapped atomic BECs have been 
computed and compared with experiments 
\cite{popov_exp,gener1,IM_ex,reidl,bergeman}. 
For temperatures $T\lesssim T_{\sc bec} / 2$, the predictions
of the PA, G1 and G2 for the lowest excitation frequencies 
differ only a few percent \cite{gener1,gener2}. 
For higher temperatures, the differences
are larger and exceed the experimental uncertainty estimates for 
measurements, but
none of the theories agrees satisfyingly with experiments \cite{gener1}. 
However, in this temperature range the dynamics of 
the thermal gas component, which these approximations do not 
take into account, is expected to have an increasingly important 
influence on the excitation eigenmodes. Consequently,
results for irrotational condensates remain somewhat
inconclusive in determining the validity of these theories.

Recently, vortex states in dilute atomic BECs have been
experimentally  realized \cite{vortex_exp}. Furthermore, by observing
the precession of vortices, the energy of the lowest
excitation, the so-called lowest core localized state (LCLS),
has been measured \cite{new_exp}. Interestingly enough, the experimental 
results for this energy agree well with the Bogoliubov
approximation \cite{prec_lcls}, while they definitely disagree with 
the picture given by the self-consistent mean-field theories: 
the latter predict the
energy of the precession mode to be positive w.r.t.\ the condensate 
state \cite{IM_stab2,Our_letter}, but experiments imply negative energies. 
We suggest that this puzzling fact could be due to incomplete 
thermalization of the (moving) vortex and/or the 
limitations of the quasi-stationary, i.e., adiabatic HFB formalism
in describing time-dependent phenomena. The adiabatic approximation
essentially fails if the kinetic rates of the system exceed frequency
separations of the excitations. Lately, we have shown that the
requirement of adiabaticity leads to a criterion for the velocity
of the moving vortex, which is violated in the precession observations
so far \cite{to_be_published}. However, if the precession 
radii---and thus the velocities---of the vortices could be reduced, 
or the physical 
parameter values appropriately adjusted in order for the system to better 
fulfill the criteria for adiabaticity and thermalization, 
one should be able to meaningfully compare experimental 
data with the predictions of the self-consistent 
equilibrium theories for the vortex states.

In this paper, we present
results of computations for the structures and excitation frequencies
of vortex states within the G1 and the G2 approximations, and compare them
with the previously computed predictions of the Popov approximation 
\cite{IM_stab2,Our_letter}. 
Outside the vortex core region, the density profiles for the condensate 
and the thermal gas component are found to differ by only a few percent 
between the PA, G1 and G2. However, in the core region the 
differences are considerably larger. This is reflected in substantial 
differences in the energy of the LCLS, which is localized in the core 
region. 

The gapless HFB-type theories considered in this paper 
can be expressed in the form of
the generalized Gross-Pitaevskii (GP) equation \cite{griffin_form,gener2}
\begin{equation}
\label{eq:GP}
[{\mathcal H}_0(\r)+U_{\rm c}(\r)|\phi(\r)|^2+2U_{\rm e}(\r)\rho(\r)]\phi(\r)
=\mu\phi(\r)
\end{equation}
for the condensate wavefunction $\phi(\r)$, and the eigenvalue equations
\begin{mathletters}
\label{eq:HFB}
\begin{eqnarray}
{\mathcal L}(\r)u_q(\r)+U_{\rm c}(\r)\phi^2(\r)v_q(\r)&=&E_q u_q(\r),\\
{\mathcal L}(\r)v_q(\r)+U_{\rm c}(\r){\phi^*}^2(\r)u_q(\r)&=&-E_q v_q(\r)
\end{eqnarray}
\end{mathletters}
for the quasiparticle amplitudes $u_q(\r)$, $v_q(\r)$, and 
eigenenergies $E_q$.
Above, ${\mathcal H}_0(\r)=-\hbar^2\nabla^2/2m+V_{\rm trap}(\r)$ is
the bare single-particle Hamiltonian for atoms of mass $m$ confined by a 
harmonic trapping potential 
$V_{\rm trap}(\r)=\frac{1}{2}m(\omega_{r}^2r^2+\omega_z^2 z^2)$ expressed in
cylindrical coordinates $\r=(r,\theta,z)$.
Furthermore, $\mu$ denotes the chemical potential, $\rho(\r)$ is the 
noncondensate density, and ${\mathcal L}(\r)\equiv {\mathcal H}_0(\r) - \mu + 
2U_{\rm c}(\r)|\phi(\r)|^2+2U_{\rm e}(\r)\rho(\r)$. The coupling functions
$U_{\rm c}(\r)$ and $U_{\rm e}(\r)$ specify the approximation:
In the Popov version $U_{\rm c}(\r)=U_{\rm e}(\r)\equiv g$, where the
coupling constant $g$ is related to the $s$-wave scattering length
$a$ by $g=4\pi\hbar^2 a/m$. In the G1 approximation, $U_{\rm e}(\r)\equiv g$ 
and $U_{\rm c}(\r)=g[1+\Delta(\r)/\phi^2(\r)]$, where $\Delta(\r)$ is 
the anomalous average of two Bose field operators, and in the G2,
$U_{\rm e}(\r)=U_{\rm c}(\r)=g[1+\Delta(\r)/\phi^2(\r)]$. 
Finally, the selfconsistency equations
\begin{eqnarray}
\rho(\r)&=&\sum_q [(|u_q(\r)|^2+|v_q(\r)|^2)n(E_q)+|v_q(\r)|^2],
\label{eq:self1}\\
\Delta(\r)&=&\sum_q [2u_q(\r)v_q^*(\r)n(E_q)+u_q(\r)v_q^*(\r)],
\label{eq:self2}
\end{eqnarray}
where $n(E_q)=(e^{E_q/k_{\rm B}T}-1)^{-1}$ is the Bose distribution 
function (the chemical potential is set equal to the condensate 
eigenenergy---for the parameter values used in the computations, the effect of 
this approximation is negligible), 
relate the noncondensate density and the anomalous average to the
positive-norm quasiparticle eigensolutions of equations (\ref{eq:HFB}).
The expression for the anomalous average is ultraviolet divergent,
and we renormalize it by subtracting the last, only implicitly
temperature-dependent terms from the sum of equation (\ref{eq:self2}).
This renormalization scheme neglects the difference between the
two- and the many-body $T$-matrices, but for dilute gases the
corrections are vanishingly small \cite{morgan}.

In this paper we consider axisymmetric single-quantum vortex states of
the form $\phi(\r)=\phi(r)e^{i\theta}$, thus neglecting possible bending
of the vortex \cite{bending}. 
Furthermore, we restrict to the case $\omega_z=0$, which
implies the system to be homogeneous in the axial direction.
However, the qualitative results concerning
especially the substantial relative differences
between the approximations in the lowest excitation energy are expected
to remain valid also for systems with $\omega_z>0$.

In accordance with the translational
symmetry in the axial direction, we impose periodic boundary conditions
at $z=\pm L/2$. By substituting the Ansatz
\begin{mathletters}
\label{eq:ansatz}
\begin{eqnarray}
u_q(\r)&=&u_q(r)e^{iq_z(2\pi/L)z+i(q_{\theta}+1)\theta},\\
v_q(\r)&=&v_q(r)e^{iq_z(2\pi/L)z+i(q_{\theta}-1)\theta},
\end{eqnarray}
\end{mathletters}
where $q_{\theta}$ and $q_z$ are integer quantum numbers, equations
(\ref{eq:HFB}) reduce to radial equations. We discretize them 
with a high-order finite-difference method and solve the
consequent matrix eigenvalue problem by using an Arnoldi
method implemented in the {\sc arpack} 
(http://www.caam.rice.edu/software/ARPACK/) subroutine libraries 
\cite{Our_letter}. 
The nonlinear Gross-Pitaevskii equation is solved using
a real-space discretization and iterative relaxation methods. Finally,
self-consistency is achieved by iteration: Using the solutions of 
equations (\ref{eq:GP}) and (\ref{eq:HFB}) corresponding to a given 
total number of particles,
new mean-field potentials are computed from the self-consistency 
equations, and the procedure is repeated until convergence. We solve
quasiparticle states explicitly up to an energy $E_{\rm c}$, and 
compute the contribution to the mean-field potentials from the 
states above $E_{\rm c}$  by using a semiclassical local
density approximation described in reference \cite{reidl}. This method allows
one to use rather small values for $E_{\rm c}$ with excellent accuracy, thus 
essentially improving the computational efficiency. In order to stabilize 
the iteration, we use underrelaxation in updating the mean-field potentials.

To facilitate comparison with previously presented results for the
Popov approximation, the physical parameter values for the gas
and the trap were chosen to be the same as those in reference \cite{IM_stab2}.
We modelled a sodium gas with the atomic mass $m=3.81\times 10^{-26}$ kg
and the scattering length $a=2.75$ nm in a trap with the radial frequency
$\nu_r=\omega_r/2\pi=200$ Hz. The density of the gas was determined
by treating $N=2\times 10^5$ atoms per length $L=10$ $\mu$m in
the axial direction. Altogether, these values yield the condensation 
temperature $T_{\sc bec}\approx 0.8$ $\mu$K. 

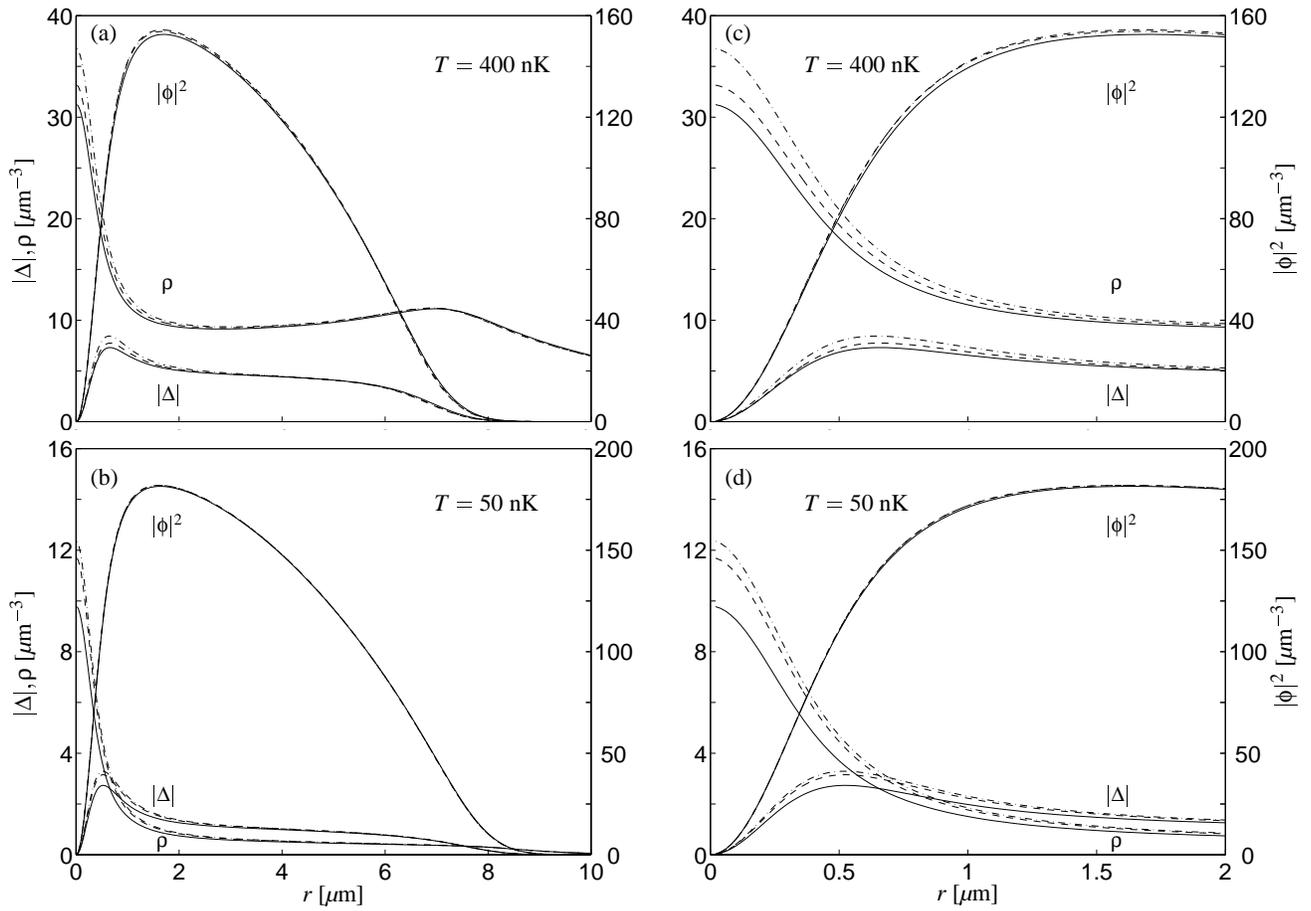
\begin{figure}
\caption{Density profiles of the vortex state for the 
condensate ($|\phi|^2$), thermal gas component ($\rho$) and
anomalous average ($|\Delta|$) in the PA (solid), G1 (dashed) and
G2 (dashed-dotted) at temperatures (a) $T=400$ nK and (b) $T=50$ nK.
Axes for the values of $\rho$ and $|\Delta|$ are on the left-hand sides,
and for $|\phi|^2$ on the right-hand sides.
Figures (c) and (d) display the vortex 
core region, where the differences in the density profiles between the
approximations are substantially larger than farther from the vortex
axis.
}
\label{fig:1}
\end{figure}

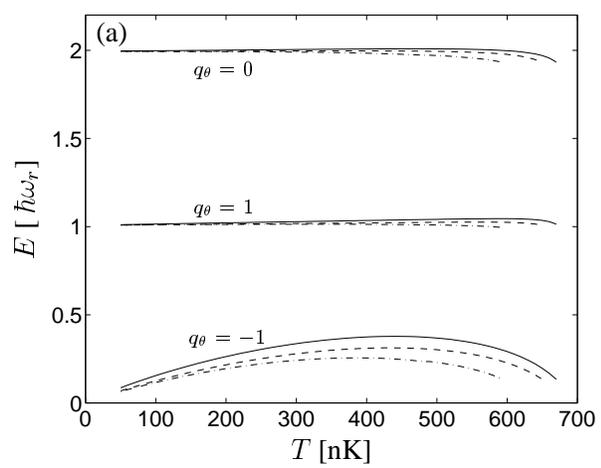
\begin{figure}
\caption{(a) Energies of the lowest excitation modes with angular momentum 
quantum numbers $q_{\theta}=-1,0,$ and $1$ in the PA (solid), 
G1 (dashed) and G2 (dashed-dotted) as functions of temperature. 
The $q_{\theta}=-1$ state is the so-called lowest core localized state
(LCLS), and the $q_{\theta}=1$ excitation is the Kohn mode.
(b) Temperature dependence of the energy of the LCLS 
within the PA, G1 and G2. Note the substantial relative
differences between the theories in this energy.
}
\label{fig:2}
\end{figure}

Figures \ref{fig:1} and \ref{fig:2} 
present results of our computations for axisymmetric
single-quantum vortex states. The density profiles for the condensate,
the noncondensate and the anomalous average are displayed in figure 
\ref{fig:1} at temperatures $T=50$ nK and $400$ nK. Outside the vortex
core region, the differences in the density profiles between the 
PA, G1 and G2 are at most a few percent at temperatures 
$T\lesssim T_{\sc bec} / 2$. In the core region, however, 
the differences are considerably larger. 
The total density of the gas is approximately 20\% larger on 
the vortex axis in the G2 than within the PA. This squeezing behavior is 
associated to the ``softening'' of the repulsive effective 
interaction in the core region due to many-body effects \cite{gener2}. 

The differences in the core densities between the various
approximations also suggest differences in the energies of the
quasiparticle excitations localized in the core region. 
Figure \ref{fig:2} displays the
energies of three such states, the lowest excitations with angular momentum 
quantum numbers $q_{\theta}=-1, 0$, and $1$, as functions
of temperature. 
For the G1 and the G2, the increased core densities are compensated by
smaller effective couplings, and the shifts in the excitation
energies are generally only a few percent for temperatures
$T\lesssim T_{\sc bec} / 2$; at higher temperatures the softening
effect of the interaction becomes more pronounced \cite{gener2,bergeman}, 
also increasing the shifts in the excitations. 
However, the lowest Kelvin mode (consisting
of the lowest $q_{\theta}=-1$ excitations) state, the LCLS, 
is especially sensitive to the structure of the core region.  
The differences in the energies 
of the LCLS between the approximations are 25--40\%  even at 
temperatures for which the predictions of the PA for the 
excitation frequencies of irrotational condensates differ by less
than 5\% from the experimental data. In addition, the temperature 
dependence of the LCLS is found to be much stronger than for the 
other states. In fact, the
lowest excitation energy vanishes in the zero-temperature limit
for all the approximations \cite{Our_letter}; the remainder of the spectrum
is essentially temperature-independent, except in the vicinity of 
$T_{\sc bec}$. The state displayed in 
figure \ref{fig:2} with $q_{\theta}=1$ is the Kohn mode, 
which should have the exact energy $E=\hbar\omega_r$ according to 
Kohn's theorem for parabolic confinement \cite{kohn}. 
Kohn's theorem is satisfied to an accuracy of 
1--6\% for all the approximations, suggesting that dynamical 
effects of the thermal gas component are small in the 
temperature range studied.

In conclusion, we  argue that future 
measurements of the lowest excitation frequencies of the vortex states 
could provide stringent tests for the validity of the mean-field
theories considered. Especially, they could be used to 
estimate the degree to which the approximations for the many-body $T$-matrices 
based on the homogeneous limit remain valid for highly inhomogeneous systems.

We thank the Center for Scientific Computing for computer resources,
and the Academy of Finland and the Graduate School in Technical Physics
for support.

%\enlargethispage*{1000mm}

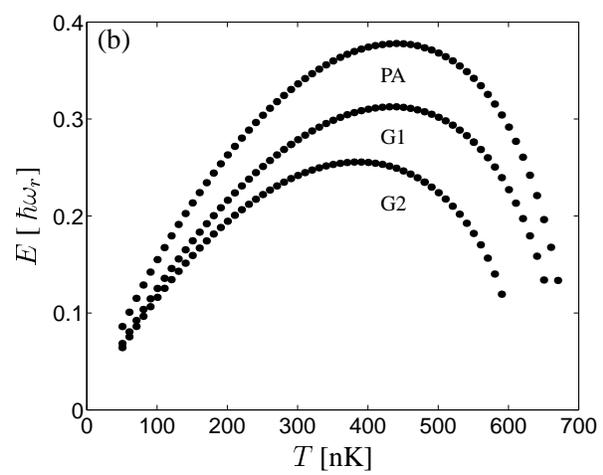
\end{document}